\documentclass[usenatbib,onecolumn]{mnras}
\usepackage{amsmath,amssymb,graphicx,paralist,tikz,pgfplots}

\newcommand*{\Mg}{1.3e11}
\newcommand*{\MG}{1.3\times 10^{11}}

\title{Scalar--tensor--vector--gravity and NGC-1277}

\author[John W. Moffat and Viktor T. Toth]{
John W. Moffat$^{1,2}$, Viktor T. Toth$^3$\\
$^1$Perimeter Institute for Theoretical Physics, Waterloo, Ontario N2L 2Y5, Canada\\
$^2$Department of Physics and Astronomy, University of Waterloo, Waterloo, Ontario N2L 3G1, Canada\\
$^3$Ottawa, Ontario K1N 9H5, Canada}

\date{Accepted XXX. Received YYY; in original form ZZZ}

\pubyear{2023}

\begin{document}

\label{firstpage}
\pagerange{\pageref{firstpage}--\pageref{lastpage}}

\maketitle

\begin{abstract}
NGC1277 is a compact but massive lenticular galaxy that shows no signs of the presence of dark matter. We find that this galaxy's behavior is consistent not only with Newtonian dynamics, but also with the predictions of Scalar--Tensor--Vector--Gravity, also known as MOG (MOdified Gravity). The compact size of the galaxy, in combination with its large mass, ensures that there are no observable deviations between the predictions of Newtonian and MOG orbital velocities within the galaxy's visible radius.
\end{abstract}


\section{Introduction}

Newly published data \citep{NGC1277:2023A} on the small lenticular galaxy NGC-1277 suggest that the behavior of this galaxy is determined entirely by Newtonian dynamics. The galaxy, which also hosts a large supermassive black hole, appears to contain no detectable amounts of dark matter. This has been presented as evidence against modified theories of gravity. Whereas putative mechanisms exist that may strip a galaxy of its dark matter content (or so the argument goes), modified gravity should result in uniform behavior: it cannot be turned off on a whim.

Scalar--Tensor--Vector--Gravity (STVG) \citep{Moffat2006a}, also known by the acronym MOG (MOdified Gravity), is one such theory. In this theory, attractive gravity is mediated by a tensor field just as in Einstein's general relativity. However, the gravitational attraction is governed by a gravitational constant $G>G_N$, where $G_N$ is Newton's constant of gravitation. This excess gravity is then canceled at short range by a Yukawa-type repulsive force that is mediated by a vector field. An additional scalar field determines the strength of the theory's gravitational constant. The range of the Yukawa-type field is also variable. For compact sources, a simple formulation exists that connects approximate values of these two parameters to the mass of the source \citep{Moffat2007e}.

This theory has been used in the past successfully, to account, for instance, for galaxy cluster dynamics \citep{Brownstein2007,MoffatRahvar2014,Israel2016}, galaxy rotation curves \citep{MoffatRahvar2013,Davari2020} including dwarf galaxies \citep{MOG2019a}, globular clusters \citep{Moffat2007a} as well as cosmology \citep{Moffat2013a,Davari2021,Moffat2021a}.

In the present paper, we employ the formalism first established in \citep{Moffat2007e} to investigate the expected behavior of NGC-1277 under MOG/STVG. After a brief introduction to the MOG theory in Section~\ref{sec:MOG}, we apply the theory to NGC-1277 in Section~\ref{sec:app}. Our conclusions are presented in Section~\ref{sec:end}.

\section{The MOG theory}
\label{sec:MOG}

Scalar--Tensor--Vector Gravity is a modified theory of gravitation (MOG) based on an action principle. In addition to the Einstein--Hilbert term, the theory postulates the existence of a vector field that results in a repulsive force of finite range, which partially cancels out attractive gravitation. The gravitational constant, the coupling strength of the vector field and the range are controlled by two additional scalar degrees of freedom.

Specifically, the theory's action integral is given by

\begin{align}
{\cal L}_G&{}=-\frac{1}{16\pi G}(R+2\Lambda),\\
{\cal L}_\phi&{}=-\frac{1}{4\pi}\omega\left[\frac{1}{4}B^{\mu\nu}B_{\mu\nu}-\frac{1}{2}\mu^2\phi^\mu\phi_\mu+V_\phi(\phi)\right],\\
{\cal L}_S&{}=-\frac{1}{G}\left[\frac{1}{2}g^{\mu\nu}\left(\frac{\nabla_\mu G\nabla_\nu G}{G^2}+\frac{\nabla_\mu\mu\nabla_\nu\mu}{\mu^2}\right)+\frac{V_G(G)}{G^2}+\frac{V_\mu(\mu)}{\mu^2}\right],\\
S&=\int d^4x\sqrt{-g}({\cal L}_G+{\cal L}_\phi+{\cal L}_S+{\cal L}_M),
\end{align}
with the matter stress-energy tensor calculated, as usual, as
\begin{align}
T_{\mu\nu}&{}=-\frac{2}{\sqrt{-g}}\frac{\delta S_M}{\delta g^{\mu\nu}},
\end{align}
where $S_M$ is the matter Lagrangian. The current associated with the vector field, in turn, is given by
\begin{align}
J^\nu&{}=-\frac{1}{\sqrt{-g}}\frac{\delta S_M}{\delta\phi_\nu}.
\end{align}

Although the theory allows for the introduction of nontrivial self-interaction potentials, usually we assume that $V_G=V_\phi=V_\mu=0$, and set the coupling constant to unity, $\omega=1$, simplifying the action. The resulting field equations are obviously more complex than in the case of general relativity, but solutions, at least in the form of approximate or numerical form, do exist. We have, in the past, successfully applied the MOG theory to reproduce important cosmological observations, for instance.

However, the main motivation behind this formalism is to create a robust theory that can reproduce large-scale gravitational dynamics (galaxy rotation, the dynamics of galaxy clusters, cosmological structure formation) without resorting to postulating a dark matter of unknown nature or origin.

A key result of the MOG theory is its acceleration law in the weak field Newtonian limit. The effective gravitational potential in the presence of a compact source is given by the equation
\begin{align}
U&{}=-\left(1+\alpha-\alpha e^{-r/r_0}\right)\frac{G_NM}{r},
\end{align}
where $G$ is the value of the $G$ scalar field in the vicinity of the source, serving as the MOG gravitational constant, $G_N$ is Newton's constant, $\alpha=(G-G_N)/G_N$ characterizes the strength of the gravitational field, while $r_0=\mu^{-1}$ is the range of the MOG repulsive force, computed as the reciprocal of the $\mu$ scalar field in the vicinity of the source.

This potential yields the acceleration law
\begin{align}
\frac{d^2{\mathbf{r}}}{dt^2}&{}=-\left[1+\alpha-\alpha \left(1+\frac{r}{r_0}\right)e^{-r/r_0}\right]\frac{G_NM}{r^3}{\mathbf{r}}.
\end{align}

As mentioned in the Introduction, this acceleration law has been used extensively to study the gravitational dynamics of systems ranging from globular clusters, through dwarf and regular galaxies, up to and beyond galaxy clusters.

\begin{figure}
\centering{\pgfplotsset{width=4.8in,height=3.2in,grid style={dotted,gray},compat=newest,
/pgf/declare function={M0=\Mg;D=6250;E=25000;a0=19;G=6.674e-11;Msol=1.98e30;kpc=3.09e19;},
/pgf/declare function={a(\M)=(a0*(\M)/(sqrt(\M)+E)^2);},
/pgf/declare function={mu(\M)=(D/sqrt(\M));},
/pgf/declare function={ar(\M,\r)=(a(\M)+1-a(\M)*(1+mu(\M)*\r)*exp(-mu(\M)*\r));},
/pgf/declare function={v(\r)=ifthenelse(abs(\r)>2,sqrt(G*Msol*M0/kpc/abs(\r))*\r/abs(\r),sqrt(G*Msol*M0/kpc/2)*\r/2)/1000*(586-110)/(883-97);},
/pgf/declare function={w(\r)=(0.5*v(\r+0.5)+0.5*v(\r+0.375)+0.25*v(\r+0.25)+0.25*v(\r+0.125)+0.25*v(\r-0.125)+0.25*v(\r-0.25)+0.5*v(\r+0.375)+0.5*v(\r-0.5)+v(\r))/4;},
/pgf/declare function={m(\r)=ar(M0,\r)*w(\r);},
}
\begin{tikzpicture}
\begin{axis}[
  xmin=-8,
  xmax=8,
  ymin=-400,
  ymax=400,
  xlabel={$r$ (kpc)},
  ylabel={$v$ (km/s)},
  grid=both,
  clip marker paths=true
]
\addplot[color=blue!40,mark=*,mark options={fill=blue!40}] table[x expr=(\thisrowno{0}-346)/(1089-346)*(10-(-5))+(-5),
                                                         y expr=(\thisrowno{1}-625)/(219-625)*(200-(-200))+(-200), row sep=crcr]{
315	236\\
379	199\\
438	170\\
501	185\\
558	278\\
624	537\\
685	659\\
745	674\\
807	635\\
864	636\\
};
\addplot[color=red!40,mark=*,mark options={fill=red!40}] table[x expr=(\thisrowno{0}-346)/(1089-346)*(10-(-5))+(-5),
                                                         y expr=(\thisrowno{1}-625)/(219-625)*(200-(-200))+(-200), row sep=crcr]{
372	148\\
427	137\\
457	132\\
479	134\\
496	128\\
508	137\\
517	141\\
525	158\\
534	171\\
543	172\\
551	197\\
560	212\\
569	226\\
577	255\\
586	331\\
593	436\\
602	524\\
611	589\\
619	621\\
627	642\\
636	658\\
645	666\\
654	687\\
662	699\\
670	696\\
679	702\\
692	711\\
709	715\\
731	706\\
760	708\\
816	689\\
};
\addplot[green!30!black,smooth,dashed,thick,domain=-7:7] (x,{-w(x)});
\addplot[black,smooth,very thick,domain=-7:7] (x,{-m(x)});
\end{axis}
\end{tikzpicture}\par}
\caption{\label{fig:MOGv}Comparison of the rotation curve predicted by MOG (solid black line) or Newtonian gravity (dashed green line), using a very simple mass model of $M=\MG$ solar masses concentrated within the inner 2~kpc, against the data presented in
 \citep{NGC1277:2023A}
(blue dots) as well as
 \citep{NGC1277:2017A}
(red dots).}
\end{figure}
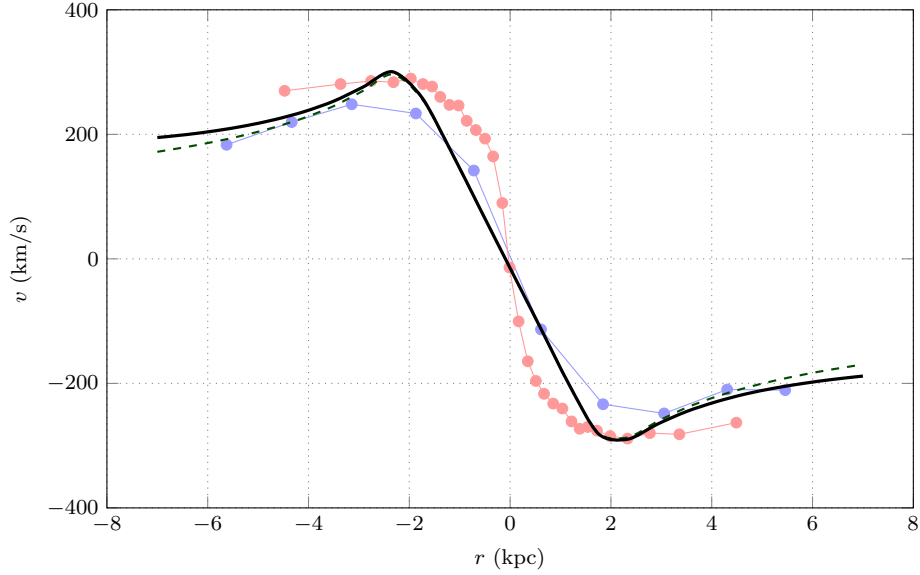

In particular, we established a set of robust semi-analytical relationships that allow us to determine, at least approximately, the values of $G$ and $\mu$ for a compact source in the weak field limit. These are given by
\begin{align}
\alpha&{}=\frac{\alpha_\infty M}{(\sqrt{M}+E)^2},\label{eq:alpha}\\
r_0&{}=\frac{\sqrt{M}}{D},\label{eq:mu}
\end{align}
with approximate values for the constants $D\simeq 6250 M_\odot^{1/2}{\rm kpc}^{-1}$ and $E\simeq 25000 M_\odot^{1/2}$ determined from best fits using data from a range of galaxies \citep{Moffat2007e}, while $\alpha_\infty\simeq 19$ was chosen to obtain a model that is consistent with a spatially flat cosmology containing baryons only.

\section{Application to NGC-1277}
\label{sec:app}

The compact lenticular galaxy NGC-1277 has been studied extensively since its discovery in 1875. It is believed to harbor a fairly large supermassive black hole with $M_{\tt SMBH}={\cal O}(10^{9}) M_\odot$. Significantly, the dynamics of NGC-1277 suggest that this galaxy is entirely or almost entirely devoid of dark matter.

As such, NGC-1277 may be viewed as evidence in favor of dark matter over alternative interpretations of galactic dynamics, notably modified theories of gravitation. The argument seems straightforward: Whereas the dark matter content of a galaxy may vary depending on the system's history, modified gravity is supposed to be a fundamental law of nature that would apply to all self-gravitating systems.

On the other hand, NGC-1277 has the distinction of being a ``relic'' galaxy that may not have had close interactions with other galaxies since its formation roughly 12 billion years ago. This presents a serious challenge to any proposed mechanism that might have stripped NGC-1277 of its dark matter content.

For these reasons, NGC-1277 is an obvious test case for both modified theories of gravity and the standard $\Lambda$-CDM cosmology with its proposed dark matter constituent.

Given the formalism we presented above, applying MOG to NGC-1277 is straightforward. The fact that NGC-1277 is compact, with much of its mass concentrated in its central region, increases our confidence that our previously developed formalism is applicable here.

The mass of NGC-1277 (with no dark matter) is estimated at $M\simeq \MG M_\odot$. Using this value, we can compute $\alpha$ and $r_0$:
\begin{align}
\alpha^{\tt NGC1277}&{}=16.61,\\
r_0^{\tt NGC1277}&{}=57.69~{\rm kpc}.
\end{align}

The dynamics of NGC-1277 were investigated to a radial distance of up to $r=6$~kpc in \citep{NGC1277:2023A}. At this distance, the gravitational acceleration predicted by the MOG theory will exceed the Newtonian prediction by the factor
\begin{align}
\frac{a_{\tt MOG}}{a_N}=1+\alpha-\alpha \left(1+\frac{r}{r_0}\right)e^{-r/r_0}\simeq 1.062.
\end{align}
Corresponding circular velocities and velocity dispersions will increase in proportion to the square root of this ratio, or about 3\%, which means that the MOG and Newtonian predictions are effectively indistinguishable.

This is demonstrated in Figure~\ref{fig:MOGv}, which shows the Newtonian/MOG prediction in the light of the rotational velocity data from \cite{NGC1277:2023A,NGC1277:2017A}.

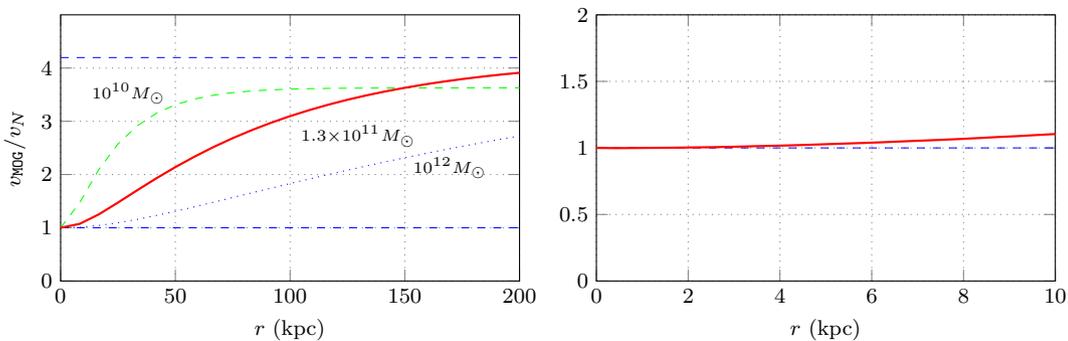
\begin{figure}
\centering{\pgfplotsset{width=3.0in,height=2.0in,grid style={dotted,gray},compat=newest,
/pgf/declare function={a(\M)=(19*(\M)/(sqrt(\M)+25000)^2);},
/pgf/declare function={mu(\M)=(6250/sqrt(\M));},
/pgf/declare function={ar(\M,\r)=(a(\M)+1-a(\M)*(1+mu(\M)*\r)*exp(-mu(\M)*\r));},
/pgf/declare function={M0=\Mg;},
}
\begin{tikzpicture}
\begin{axis}[
  xmin=0,
  xmax=200,
  ymin=0,
  ymax=5,
  xlabel={$r$ (kpc)},
  ylabel={$v_{\tt MOG}/v_N$},
  grid=both,
  ytick={0,1,2,3,4},
]
\addplot[blue,dashed,line width=0.25pt,domain=0:200] (x,{1});
\addplot[blue,dashed,line width=0.25pt,domain=0:200] (x,{sqrt(a(M0)+1)});
\addplot[blue,thin,dotted,domain=0:200] (x,{sqrt(ar(1e12,x))});
\addplot[green,thin,dashed,domain=0:200] (x,{sqrt(ar(1e10,x))});
\addplot[red,thick,domain=0:200] (x,{sqrt(ar(M0,x))});
\node[] at (axis cs: 30,3.5) {$^{10^{10}M_\odot}$};
\node[] at (axis cs: 130,2.7) {$^{\MG M_\odot}$};
\node[] at (axis cs: 170,2.1) {$^{10^{12}M_\odot}$};
\end{axis}
\end{tikzpicture}
\begin{tikzpicture}
\begin{axis}[
  xmin=0,
  xmax=10,
  ymin=0,
  ymax=2,
  xlabel={$r$ (kpc)},
  grid=both,
]
\addplot[blue,dashed,line width=0.25pt,domain=0:10] (x,{1});
\addplot[red,thick,domain=0:10] (x,{sqrt(ar(M0,x)});
\end{axis}
\end{tikzpicture}\par}
\caption{\label{fig:plots}Plots showing the anticipated increase in rotational velocities or velocity dispersions in the MOG theory, in comparison with Newtonian physics. Plotted is the ratio $\sqrt{a_{\tt MOG}/a_N}$. Left: the horizontal range is extended to 200~kpc, showing a transitional region dominated by a Tully-Fisher type behavior, extending on the right to the region where the repulsive vector force wanes, and the Keplerian rotation curves are re-established but with a rotational velocity substantially greater than the Newtonian value. Along with the $M=\MG M_\odot$ result (solid red line) two additional representative cases are also shown. Right: within the innermost 10~kpc, the behavior is indistinguishable from the Newtonian prediction. Thin dotted lines mark the minimum (1) and asymptotic maximum ($\sqrt{1+\alpha}$) values for $M=\MG M_\odot$.}
\end{figure}

One may wonder why there is such a small deviation between the predictions of Newtonian gravity and the MOG theory. This is best explained through Figure~\ref{fig:plots}, which shows how the MOG theory alters the velocities of circular orbits for objects of varying mass. For a heavier object, the overall effect will be larger, in accordance with (\ref{eq:alpha}). However, for a heavier source, the mass associated with the repulsive vector field of the theory is smaller, and correspondingly, the range increases (\ref{eq:mu}): this means that the behavior of the galaxy remains Newtonian up to a larger radius. In the specific case of NGC-1277, the relatively large stellar mass of the galaxy, in combination with its compact size, is what results in near-Newtonian behavior in MOG. To paraphrase a statement from \citep{NGC1277:2023A}, originally made about the Milgromian theory, ``[...] a radius $R = 13$~kpc should be explored to be able to probe the fully [MOG] regime. This is about twice the radius that we cover [...]''. Therefore, it should come as no surprise that there is no discernible difference between the MOG and Newtonian predictions.

\section{Discussion}
\label{sec:end}

NGC-1277 is a unique galaxy. It is considered a ``relic'' galaxy, with indications that it did not have significant gravitational interactions with other galaxies since its formation. It has considerable stellar mass but it is also very compact. This galaxy, therefore, offers a unique opportunity to study gravitational dynamics.

Observations of NGC-1277 suggest that the galaxy's dynamics can be entirely modeled using Newtonian dynamics. This has been interpreted as evidence that the galaxy contains no significant quantities of dark matter. It may also be viewed as evidence against modified theories of gravitation.

Indeed, we confirm that the dynamics of NGC-1277 are easily modeled using Newtonian physics and only the galaxy's observed stellar mass. However, the behavior of NGC-1277 is also consistent with Scalar--Tensor--Vector gravity, STVG/MOG.

The secret lies in the compact nature of the galaxy. Although it is massive, with a stellar mass in excess of $M_\star\gtrsim 1.3\times 10^{11}M_\odot$, the galaxy has a visible radius less than $\sim 8$~kpc. In the MOG theory, the effective gravitational acceleration arises as a combination of two factors: a variable gravitational coupling parameter governed by a scalar field, and a massive vector field that provides a repulsive force of limited range. This ensures that sufficiently compact objects in the weak field limit follow near-Newtonian dynamics (different considerations apply to objects with strong gravity, such as black holes \citep{Moffat2015a}). The scalar field that controls the gravitational coupling parameter, as well as the range of the vector field can be derived using semianalytical formulae from the source mass.

When we apply this formalism to NGC-1277, we find no discernible difference between the MOG prediction and Newtonian physics. Thus, far from presenting a challenge, NGC-1277 may be seen as a validation of MOG. We note that this agreement with the data is achieved with no additional assumptions, no adjustment of any of the theory's parameters: the values used are the same values that, over the past 15+ years, have been used successfully to match the properties and dynamics of a range of galaxies and other objects.

It may appear striking that MOG predicts no discernible deviation from Newtonian gravity at 6~kpc for NGC-1277. This is due to a combination of two factors: this galaxy is both compact and massive. Its compact nature justifies the use of the expressions (\ref{eq:alpha})--(\ref{eq:mu}), applicable to point or compact sources; the large mass ensures that $r_0$ is large.

Looking at past studies, we investigated a few galaxies of similar or even greater mass, e.g., NGC-801, NGC-5533 or NGC-6674 \citep{Brownstein2006a}. The observed rotation curves of these galaxies extend to several 10~kpc, with useful data only starting at $\sim 6$~kpc. To the extent that any deviation from the Newtonian prediction exists at such low radii, it appears that it is a result of the fact that these galaxies are substantially less compact than NGC-1277 and consequently, at small radii $\alpha$ and $r_0$ will be variable quantities not well predicted by (\ref{eq:alpha})--(\ref{eq:mu}). At larger radii, the behavior of these galaxies is also consistent with the MOG prediction for compact sources.

Lastly, we mention that another galaxy, NGC-1278, was also studied by \cite{NGC1277:2023A}. However, for this galaxy no clean rotation curve is offered, and therefore we made no attempt to study it. We note nonetheless that NGC-1278 does not appear to exhibit any unusual behavior and therefore, we expect it to show characteristics that are consistent with the MOG theory.

\section*{Acknowledgments}

We thank Martin Green for valuable discussions. Research at the Perimeter Institute for Theoretical Physics is supported by the Government of Canada through industry Canada and by the Province of Ontario through the Ministry of Research and Innovation (MRI). VTT acknowledges the generous support of David H. Silver, Plamen Vasilev and other Patreon patrons.

\section*{Data Availability}

No data was generated and/or analysed to produce this article.

\bibliographystyle{mnras}
\bibliography{refs}

\label{lastpage}

\end{document}